\documentclass[20pt]{article}

\usepackage[preprint]{neurips_2023}

\usepackage[utf8]{inputenc} % allow utf-8 input
\usepackage[T1]{fontenc}    % use 8-bit T1 fonts
\usepackage{hyperref}       % hyperlinks
\usepackage{url}            % simple URL typesetting
\usepackage{booktabs}       % professional-quality tables
\usepackage{amsfonts}       % blackboard math symbols
\usepackage{nicefrac}       % compact symbols for 1/2, etc.
\usepackage{microtype}      % microtypography
\usepackage{xcolor}         % colors
\usepackage{graphicx}
\usepackage{subcaption}
\usepackage{multirow}
\usepackage{array}
\usepackage{rotating}
\usepackage{wrapfig}
\usepackage{amsmath}
\usepackage{amssymb}

\title{ProPath: Disease-Specific \textsc{Pro}tein Language Model for Variant \textsc{Path}ogenicity}

\author{%
  Huixin Zhan\\
  Division of Artificial Intelligence in Medicine\\
  Cedars-Sinai Medical Center\\
  Los Angeles, CA, USA, 90048 \\
  \texttt{Huixin.Zhan@cshs.org} \\
  \And
  Zijun (Frank) Zhang\thanks{Corresponding author.} \\
  Division of Artificial Intelligence in Medicine\\
  Cedars-Sinai Medical Center\\
  Los Angeles, CA, USA, 90048 \\
  \texttt{zijun.zhang@cshs.org} \\
}

\begin{document}

\maketitle

\begin{abstract}
Clinical variant classification of pathogenic versus benign genetic variants remains a pivotal challenge in clinical genetics. Recently, the proposition of protein language models has improved the generic variant effect prediction (VEP) accuracy via weakly-supervised or unsupervised training. However, these VEPs are not disease-specific, limiting their adaptation at point-of-care. To address this problem, we propose a disease-specific \textsc{pro}tein language model for variant \textsc{path}ogenicity, termed ProPath, to capture the pseudo-log-likelihood ratio in rare missense variants through a siamese network. We evaluate the performance of ProPath against pre-trained language models, using clinical variant sets in inherited cardiomyopathies and arrhythmias that were not seen during training. Our results demonstrate that ProPath surpasses the pre-trained ESM1b with an over $5\%$ improvement in AUC across both datasets. Furthermore, our model achieved the highest performances across all baselines for both datasets. Thus, our ProPath offers a potent disease-specific variant effect prediction, particularly valuable for disease associations and clinical applicability.
\end{abstract}

\section{Introduction}

Clinical variant interpretation is transforming precision medicine, yet limitations exist that prevent its further adaptations and utilities~\citep{katsanis2013molecular}.
Following a disease diagnosis, the identification and classification of pathogenic vs benign genetic variant has important clinical implications.
The outcome of clinical variant interpretation provides a basis for clinical screening~\citep{cocchi2020clinical,xie2023genomic} and genetic testing of first-degree family members~\citep{ni2023screening}, and may serve as a prognostic marker for the affected patient~\citep{lee2019cag,musunuru2020genetic}. Currently, the utility of genetic testing is limited by the fact that a substantial proportion (30-50\%) of yielded variants are classified as variant of uncertain significance (VUS) according to the ACMG guidelines~\citep{richards2015standards}. The presence of VUSs complicates the genetic counseling and patient management, while they can not be used in clinical decision making. Given the large number of genetic variations classified as VUS, both common and rare, \textit{in silico} methods are ideal avenues to aid the clinical variant interpretation, and to facilitate downstream prioritization of experimental validation and clinical testing~\citep{frazer2021disease}.
%In clinical genetics, coding variants that alter protein amino acid sequences are critical because of their clear association with diseases and therapeutic potential. 

Predicting the phenotypic outcomes of genetic variations, commonly referred to as variant effect prediction (VEP), remains challenging. This is because existing variant annotations are limited in amount and biased by human curations. As such, it is highly desirable for computational VEP methods to be trained via weakly-supervised or unsupervised approaches that are independent of human bias. Conventionally, evolutionary conservation-based methods~\citep{reva2011predicting,rentzsch2019cadd} have been considered as weak evidence (PP3/BP4) for clinical variant interpretation in ACMG guidelines~\citep{richards2015standards}. Using a more sophisticated machine learning approach, a deep generative model EVE based on variational autoencoders~\citep{frazer2021disease} achieved state-of-the-art performance in classifying clinical variants and outperformed conservation-based method in ClinVar~\citep{landrum2014clinvar}. EVE is trained on multiple sequence alignment (MSA) that captures evolutionary-related sequence variations across species, with the goal of reconstructing the MSA from a latent bottleneck. More recently, the emergence of protein language models have expanded the arsenal of unsupervised and weakly-supervised VEP methods with new, powerful tools. Similar to EVE, the MSA transformer~\citep{rao2021msa} is also trained on MSA data, but employs a masked language modeling objective using self-attention mechanisms. 
The AlphaMissense model~\citep{cheng2023accurate} leverages the potential of integrating evolutionary data with protein structural modeling. It's trained on population frequency data, utilizes sequences from MSAs, and incorporates predicted structural contexts, all of which collectively augment its predictive performance. 
Strikingly, ~\citet{brandes2023genome} proposed a zero-shot workflow that adapts ESM1b for protein sequences of any length and employed it to predict potential missense variant impacts in the human genome. 
Unlike the above methods leveraging MSAs, EMS1b-zero shot does not explicitly rely on any evolutionary data.
Breaking the reliance on MSA is significant, especially for orphan genes with poor MSA coverage~\citep{chowdhury2022single,michaud2022language} and for rare variants that are underrepresented in population~\citep{manolio2009finding}. 
The zero-shot ESM-1b outperformed EVE, suggesting that large protein language models have learned and generalized over evolutionary constraints for the VEP task. 

However, unsupervised VEP methods are not disease specific, substantially limiting their utility and adoption in clinical variant interpretations at point-of-care. This is especially problematic when different variants in a single gene would lead to various closely-related, but distinct disease phenotypes, such as in cardiomyopathies~\citep{mcnally2015genetic,zhang2021disease}. For each missense genetic variant, unsupervised VEP will yield a score representing whether the variant is damaging to the disease-relevant protein function or structure, without distinguishing important disease-specific parameters underlying the gene-disease relationship. Such gene-disease relationships include the distinction between gain-of-function (GoF) vs loss-of-function (LoF), and primary vs modifier effects, etc. For instance, GoF vs LoF may lead to distinct phenotypes under a disease-specific context. Computational modeling for disease-specific variants is difficult, because pathogenic and benign variant annotations are even more sparse when restricted to a single disease condition. Failure to account for these disease-specific information will decrease the predictive power, and more importantly, lead to incorrect clinical decisions and sub-optimal patient management.  

To address this challenge, we introduce a disease-specific \textsc{pro}tein language model for variant \textsc{path}ogenicity (ProPath), designed to more effectively capture the pseudo-log-likelihood ratio of rare missense variants in disease-specific contexts. Our ProPath is an analogy to the concept of semantic textual similarity (STS)~\citep{han2013umbc_ebiquity} in natural language processing (NLP), where a score is assigned to measure the similarity between two text segments~\citep{reimers2019sentence}. Employing a siamese network, the similarity between the wild-type sequence and the mutated sequence is determined by their embeddings from two weight-sharing protein language model branches, and the pretrained knowledge in protein language model is fine-tuned by a small set of disease-specific variant annotations.
%Employing a siamese network, the similarity between each sentence pair is determined by calculating their embedding similarity. Similarly, the wild-type sequence and its corresponding mutation are fed into our DS-SPLM. This step is performed to compute the masked language model logits, upon which the pseudo-log-likelihood ratio is calculated.
We evaluate ProPath's performance against the zero-shot performance of pre-trained language models, utilizing clinical variant sets in inherited cardiomyopathies and arrhythmias, which were unseen during training. The experimental results show that our proposed ProPath outperforms the pre-trained ESM1b with an over $5\%$ improvement in AUC across both cardiomyopathies and arrhythmias datasets. Moreover, our model achieves the highest performances among all baselines in both datasets. Consequently, our ProPath offers a potent
disease-specific VEP for disease associations, clinical applicability, and better understanding of disease mechanisms.

 Our contributions are as follows:
 \begin{itemize}
     \item Technically, we introduce a novel disease-specific protein language model for variant pathogenicity (ProPath) to more effectively capture the pseudo-log-likelihood ratio in rare missense variants using both masked language model logits through a siamese network.
     \item Clinically, we propose an efficient way to fine-tune a protein language model to estimate the probability of pathogenicity for rare missense variants in inherited cardiomyopathies and arrhythmias to address a persistent challenge in clinical genetics.
     \item Our ProPath model sets a new benchmark by obtaining the highest statistics across all baselines in both datasets, achieveing $91\%$ and $94\%$ AUPR on the cardiomyopathies and arrhthmias variant sets, respectively.
 \end{itemize}

\section{Methods}
In the context of protein language models, the ESM1b employs a Masked Language Model (MLM) setting. Within this MLM paradigm, specific amino acid residues in protein sequences are ``masked'' or hidden, and the model is trained to predict the identity of these masked residues. As the model makes these predictions, it produces raw scores or predictions for each potential amino acid that could replace the masked residue, commonly referred to as ``MLM logits''. In subsection~\ref{sec:mlm_logits} of the supplemantary material, an example of the MLM logtis for $s^{\text{WT}}$ and $s^{\text{mut}}$ are shown in Figure~\ref{fig:mlm-1} and Figure~\ref{fig:mlm-2}, respectively. The logits predicted by a protein language model for observing the input amino acid $s_i$ at position $i$ given the sequence $s$ are shown in red frames.  When passed through an activation function like softmax, these logits provide probabilities over the possible amino acids, guiding the prediction process.

\begin{figure}[!htbp]
    \centering
    \includegraphics[width=12cm,height=2.5cm]{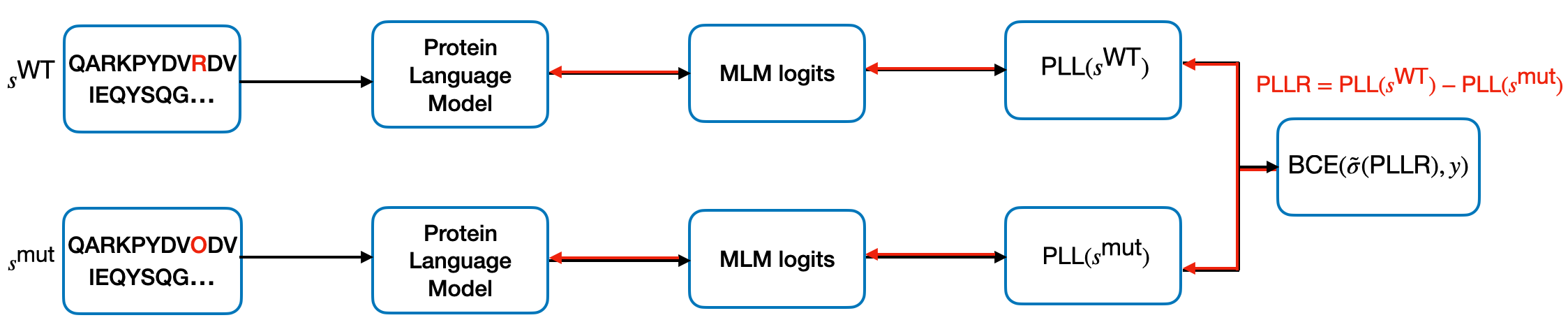}
    \caption{ProPath architecture with objective function to fine-tune on disease specific dataset. The two protein language models have tied weights (siamese network structure).}
    \label{fig:siamese_network}
\end{figure}

\paragraph{Pseudo-log-likelihood ratio computation}
In order to fine-tune ProPath on each pair of wild-type and mutant sequences, i,e., $s^{\text{WT}}$ and $s^{\text{mut}}$, we create a siamese network with two weight-sharing protein language model branches (shown in Figure~\ref{fig:siamese_network}) to update the weights such that the produced MLM logits are both semantically meaningful and can be compared via pseudo-log-likelihood ratio (PLLR). For a sequence $s = s_1, …, s_L$, the pseudo-log-likelihood is calculated as $\text{PLL}(s) = \sum_{i=1}^L \log P(x_i = s_i|s)$, where $L$ denotes the sequence length, $s_i$ represents the amino acid at position $i$, and $\log P(x_i = s_i|s)$ denotes the log-likelihood predicted by the protein language model when observing amino acid $s_i$ at position $i$ within sequence $s$. The PLLR between the $s^{\text{WT}}$ and $s^{\text{mut}}$ is then computed as:
\begin{equation}\label{eq:ab_pllr}
    \lambda = |\text{PLL}(s^{\text{WT}}) - \text{PLL}(s^{\text{mut}})|, 
\end{equation}
because a wild-type sequence typically tends to have a higher log-likelihood in a protein language model and a mutation disrupts the protein with a lower log-likelihood.
\paragraph{Classification objevtive function}
To perform the classification of pathogenic vs benign genetic variant via the siamese network, we will utilize a binary cross entropy loss. Binary cross entropy, often referred to as logarithmic loss or log loss, penalizes the model for incorrect labeling of data classes by monitoring deviations in probability during label classification. In order to fine-tune the siamese network using binary cross entropy loss, we calibrate the PLLR to a probability $\hat{\sigma}$ between $0$ to $1$ by: $\hat{\sigma}(\lambda) = 2\sigma(\lambda) - 1,$
due to the sigmoid function $\sigma$ is between $0.5$ to $1$ for |PLLR| between $0$ to $+\infty$. We then fine-tune the siamese network with the binary cross entropy loss as follows:
\begin{equation}
    \mathcal{L}_{\text{BCE}} = y \cdot \log (\hat{\sigma}(\lambda)) + (1-y)\cdot \log (1 - \hat{\sigma}(\lambda)).
\end{equation}
Thus, the objective is to maximize the PLLR to distinct the MLM logits between $s^{\text{WT}}$ and $s^{\text{mut}}$ if the mutation is pathogenic and vice versa.

\section{Results}
In this section, we present the dataset, set-up, and experimental results for both datasets. We further analyze the distribution of PLLR values and the MLM logits differences between zero-shot ESM1b and fine-tuned ProPath to show the effectiveness of our method in identifying pathogenic and benign rare missense genetic variants.
\subsection{Dataset and set-up}
In our experiments, we focus on clinical variant sets pertaining to inherited cardiomyopathies and arrhythmias. We employed a pre-compiled dataset of rare missense pathogenic and benign variants, defined by a cohort-based approach, in cardiomyopathy and arrhythmias, respectively. The details can be found in the previous report ~\citet{zhang2021disease}. The statistics for both datasets are shown in Table~\ref{tab:dataset} (subsection~\ref{sp:dataset} in the supplementary material). For more details about the dataset and details for hyperparameters, we recommend the readers refer to subsection~\ref{sp:dataset} in the supplementary material. We fine-tune our ProPath on two base protein language models, i.e., esm1b\_t33\_650M\_UR50S (ESM1b)~\citep{rives2021biological} and esm2\_t33\_650M\_UR50D (ESM2)~\citep{lin2023evolutionary}. ESM1b is a protein language model endowed with $650$ million parameters and trained on a corpus of $250$ million protein sequences spanning various organisms. Its training involved the MLM task, where certain residues from input sequences are randomly obscured, challenging the model to accurately predict the correct amino acid for each masked position. On the other hand, ESM2 the latest model in the ESM family with improvements in architectures and training parameters, and is benchmarked to outperform ESM1b at a comparable number of parameters. Over the course of training, ESM2 sees around $65$ million unique sequences. 
% Briefly, the authors collected 356 unique rare missense variants specifically associated with cardiomyopathy genes. The authors curated a set of 252 unique rare missense variants linked with established arrhythmia genes from ClinVar. For a benign set, they gathered 302 unique rare missense variants in cardiomyopathy genes and 237 in arrhythmia genes from the targeted sequencing of 2,090 healthy individuals. For training and testing set distribution, 440 and 326 cardiomyopathy variants were designated for training and testing, respectively. For the arrhythmias category, 218 variants were used for training and 166 for testing.
\subsection{Experimental results on cardiomyopathies dataset}
\paragraph{ProPath outperforms pre-trained protein language models on cardiomyopathies.}
We evaluate the ability of our ProPath to capture disease-specific information related to cardiomyopathies (CM) through fine-tuning. Figure~\ref{fig:zs-cm-m} and Figure~\ref{fig:zs-cm} (in subsection~\ref{sp:pr-zs-cm} of the supplementary material) display the receiver operating characteristic (ROC) and precision-recall (PR) curves for both pretrained ESM1b and ESM2, respectively. To derive the ROC and PR, we use two variations of PLLRs. The first uses the absolute value of PLLR as detailed in Equation~\ref{eq:ab_pllr}, while the second, termed weighted PLLR, is calculated as $\tilde{\lambda} = \text{PLL}(s^{\text{WT}})/L_{\text{WT}} - \text{PLL}(s^{\text{mut}})/L_{\text{mut}}.$ As observed from Figure~\ref{fig:zs-cm-a} and~\ref{fig:zs-cm-b}, pretrained ESM1b achieves an AUC of $0.82$ and an AUPR of $0.84$ for PLLR on CM. All PR curve and AUPR results can be found in the supplementary materials. Specifically, Figure~\ref{fig:zs-cm-b} is located in subsection~\ref{sp:pr-zs-cm} of the supplementary material. Similarly, pretrained ESM2 registers an AUC of $0.74$ and an AUPR of $0.76$ in Figure~\ref{fig:zs-cm-c} and~\ref{fig:zs-cm-d}, respectively. These results indicate that pretrained protein language models still have challenges in precisely distinguishing rare genetic variants as pathogenic or benign for disease-specific conditions such as cardiomyopathy.
%\hfill
%\begin{wrapfigure}{r}{0.4\textwidth}
\begin{figure}[!htbp]
    \centering
    \begin{subfigure}{0.45\textwidth}
        \centering
        \includegraphics[width=7.5cm,height=5cm]{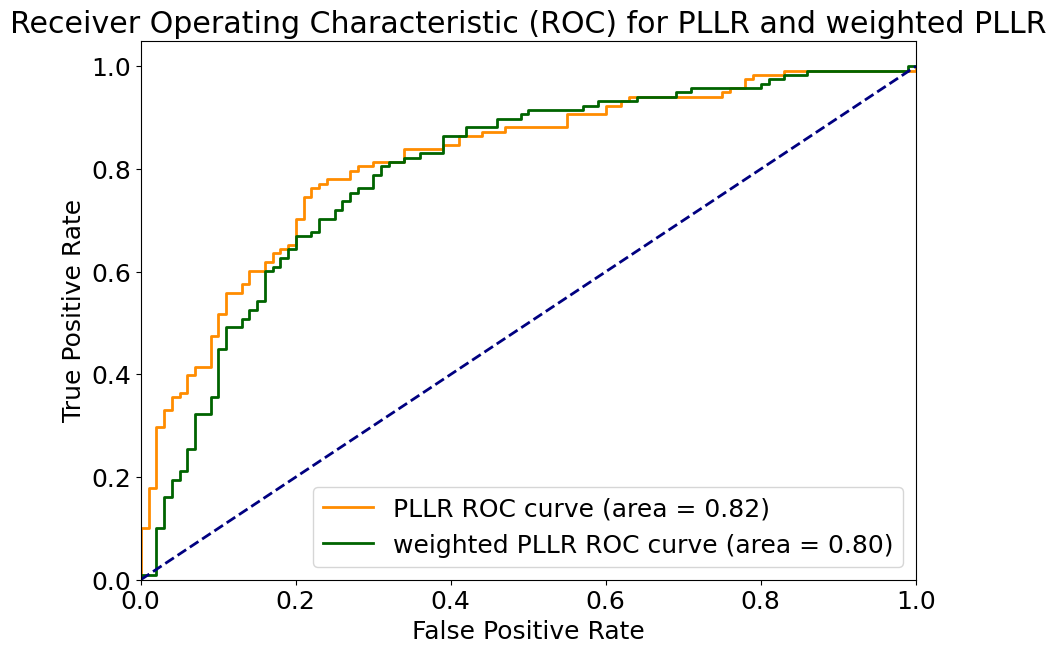}
        \caption{ROC for pretrained ESM1b on CM}
        \label{fig:zs-cm-a}
    \end{subfigure}
    \begin{subfigure}{0.45\textwidth}
        \centering
        \includegraphics[width=7.5cm,height=5cm]{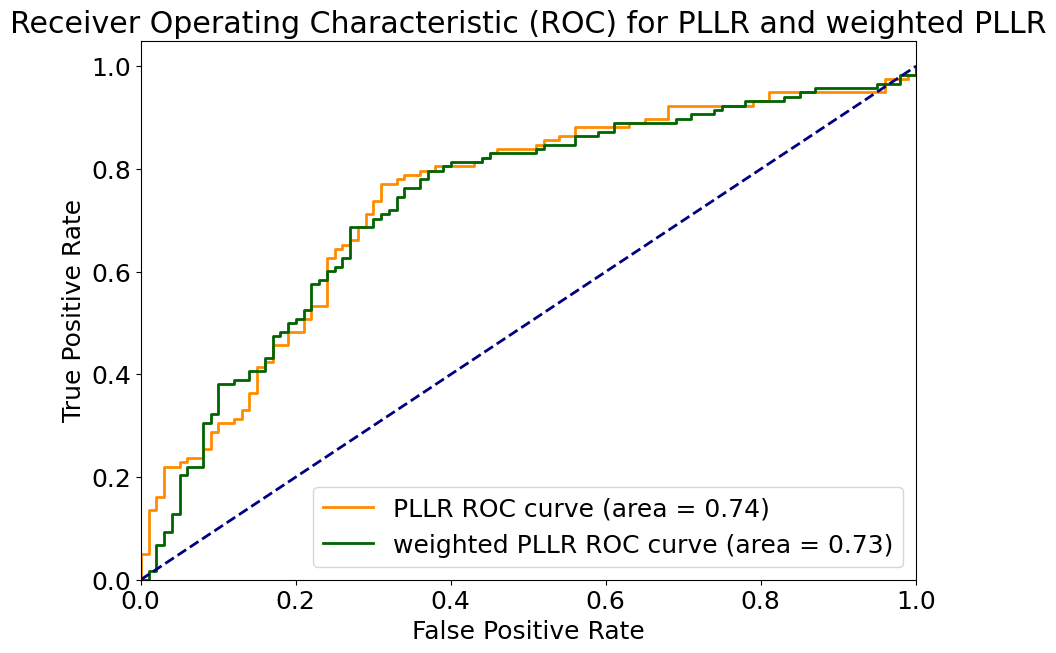}
        \caption{ROC for pretrained ESM2 on CM}
        \label{fig:zs-cm-c}
    \end{subfigure}
   \caption{Zero-shot AUC performances on CM.}
    \label{fig:zs-cm-m}
\end{figure}

\begin{figure}[!htbp]
    \centering
    \begin{subfigure}{0.45\textwidth}
        \centering
        \includegraphics[width=7.5cm,height=5cm]{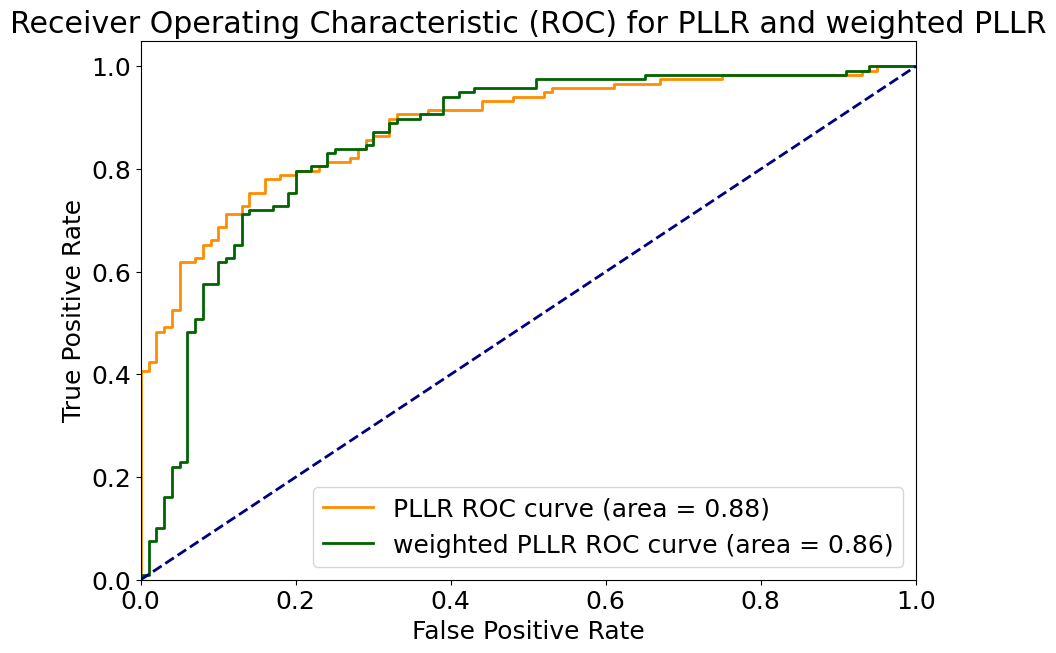}
        \caption{ROC for fine-tuned ProPath via ESM1b on CM}
        \label{fig:}
    \end{subfigure}
    \hspace{1pt}
    \begin{subfigure}{0.45\textwidth}
        \centering
        \includegraphics[width=7.5cm,height=5cm]{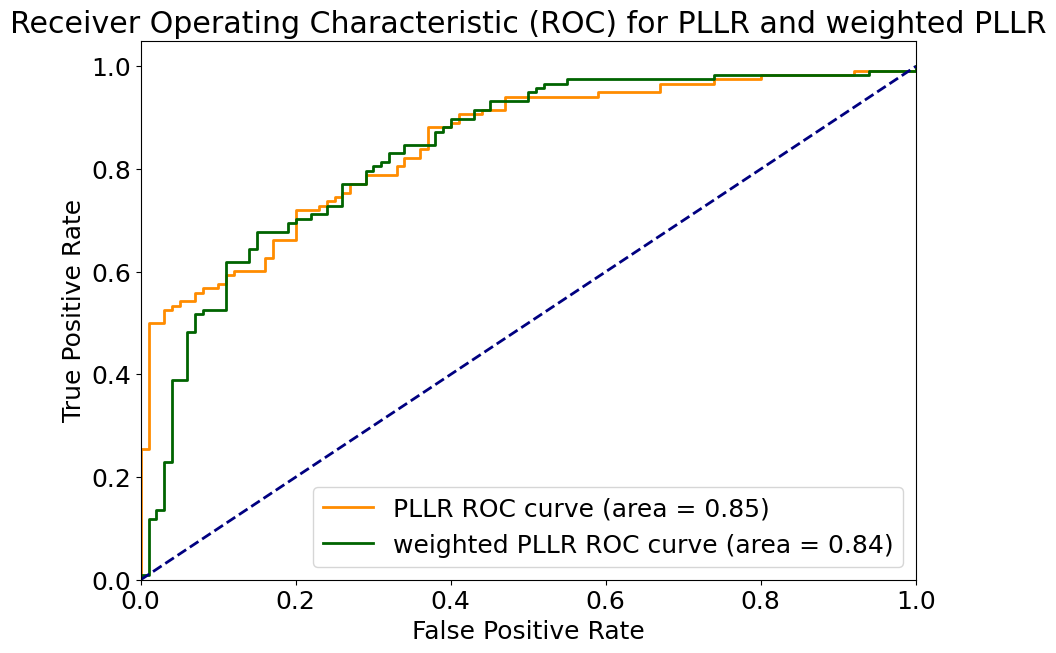}
        \caption{ROC for fine-tuned ProPath via ESM2 on CM}
        \label{fig:}
    \end{subfigure}
        \caption{Fine-tuned ProPath performances on CM.}
    \label{fig:fs-cm-m}
\end{figure}

In contrast, the fine-tuned ESM1b attains an AUC of $0.88$ for PLLR on CM, as shown in Figure~\ref{fig:fs-cm-m}. Meanwhile, its AUPR of $0.91$ is illustrated in Figure~\ref{fig:fs-cm}, which can be found in the supplementary material's subsection~\ref{sp:pr-ft-cm}. Similarly, the fine-tuned ESM2 attains an AUC of $0.85$ and an AUPR of $0.88$. These results indicate that the fine-tuned ProPath achieves a much higher precision increase in distinguishing disease-specific variants when classifying genetic variants. This demonstrates that our ProPath more effectively captures the PLLR in rare missense variants.

\paragraph{ProPath outperforms all baseline methods on CM.}
We further evaluate the ability of our ProPath to capture disease-specific information related to CM through fine-tuning via comparing with five baseline methods shown in Table~\ref{tab:cm-baselines}. The best performances in terms of AUC and AUPR are shown in bold. Our ProPath, utilizing ESM1b as its foundational model, outperforms the five baselines with an AUC of $0.88$ and an AUPR of $0.91$. Therefore, this reaffirms the capability of ProPath as a robust tool for disease-specific VEP in discerning disease associations.
%\hfill
%\begin{wraptable}{r}{10cm}
\begin{table}[!htbp]
\caption{ Performance for cardiomyopathy variant pathogenicity prediction.}
\label{tab:cm-baselines}
\centering
\scalebox{1.0}{
 \begin{tabular}{l |>{\centering\arraybackslash}p{5cm} >{\centering\arraybackslash}p{1.5cm} >{\centering\arraybackslash}p{1.5cm} >{\centering\arraybackslash}p{1.5cm} }
    \hline
     \multirow{2}{*}{\textbf{Category}}   & \multirow{2}{*}{\textbf{Algorithm}} & \textbf{Disease-specific} & \multirow{2}{*}{\textbf{AUC}} & \multirow{2}{*}{\textbf{AUPR}}\\
      \hline
    \multirow{3}{*}{Boosting models} 
         & XGBoost~\citep{zhang2021disease} & $\checkmark$ &  $0.87$ & $0.9$  \\
         & AdaBoost~\citep{zhang2021disease} & $\checkmark$ & $0.88$ & $0.9$ \\
         & M-CAP~\citep{jagadeesh2016m} & $\times$ &$0.79$ & $0.8$\\
          \hline
    \multirow{1}{*}{Ensemble models}  
         & REVEL~\citep{ioannidis2016revel} & $\times$ & $0.81$ & $0.79$ \\
         \hline
    \multirow{2}{*}{Language models}
         & ESM1b~\citep{brandes2023genome} &$\times$ &$0.82$ & $0.84$ \\
         & ProPath  & $\checkmark$ & $\mathbf{0.88}$ &$\mathbf{0.91}$ \\
         \hline
    \end{tabular}
}
\end{table}
%\end{wraptable}

\subsection{Experimental results on arrhythmias dataset}
\paragraph{ProPath outperforms pre-trained protein language models on arrhythmias.}
Next, we evaluate ProPath's capability to extract disease-specific insights related to arrhythmias (ARM). Figure~\ref{fig:zs-arm-m} and Figure~\ref{fig:zs-arm} (in subsection~\ref{sp:pr-zs-arm} of the supplementary material) show the ROC and PR curves for both pretrained ESM1b and ESM2. Figure~\ref{fig:zs-arm-a} and~\ref{fig:zs-arm-b} reveal that the pretrained ESM1b obtains an AUC of $0.88$ and an AUPR of $0.89$ for PLLR on ARM. However, using weighted PLLR, its performance increases to an AUC of $0.89$ and an AUPR of $0.91$, outperforming the regular PLLR. Similarly, pretrained ESM2 obtains an AUC of $0.82$ and an AUPR of $0.87$ using PLLR while AUC=$0.84$ and AUPR=$0.88$ using weighted PLLR in Figure~\ref{fig:zs-arm-c} and~\ref{fig:zs-arm-d}. This demonstrates that different scoring and normalization methods have distinct relative predictive powers when analyzing different diseases, consolidating the need for disease-specific fine-tune modeling. 
%This underscores that while some pretrained protein language models might perform well on specific clinical variant datasets, their adaptability and precision can waver across distinct foundational models when discerning disease-specific variants in genetic classification.

\begin{figure}[!htbp]
    \centering
    \begin{subfigure}{0.45\textwidth}
        \centering
        \includegraphics[width=7.5cm,height=5cm]{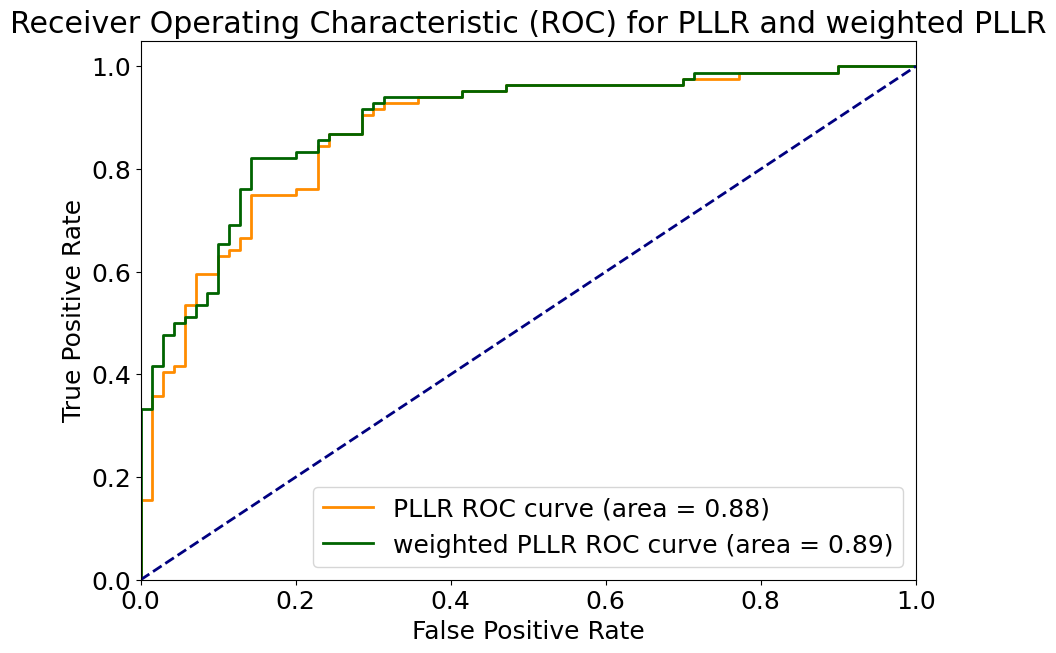}
        \caption{ROC for pretrained ESM1b on ARM}
        \label{fig:zs-arm-a}
    \end{subfigure}
    \begin{subfigure}{0.45\textwidth}
        \centering
        \includegraphics[width=7.5cm,height=5cm]{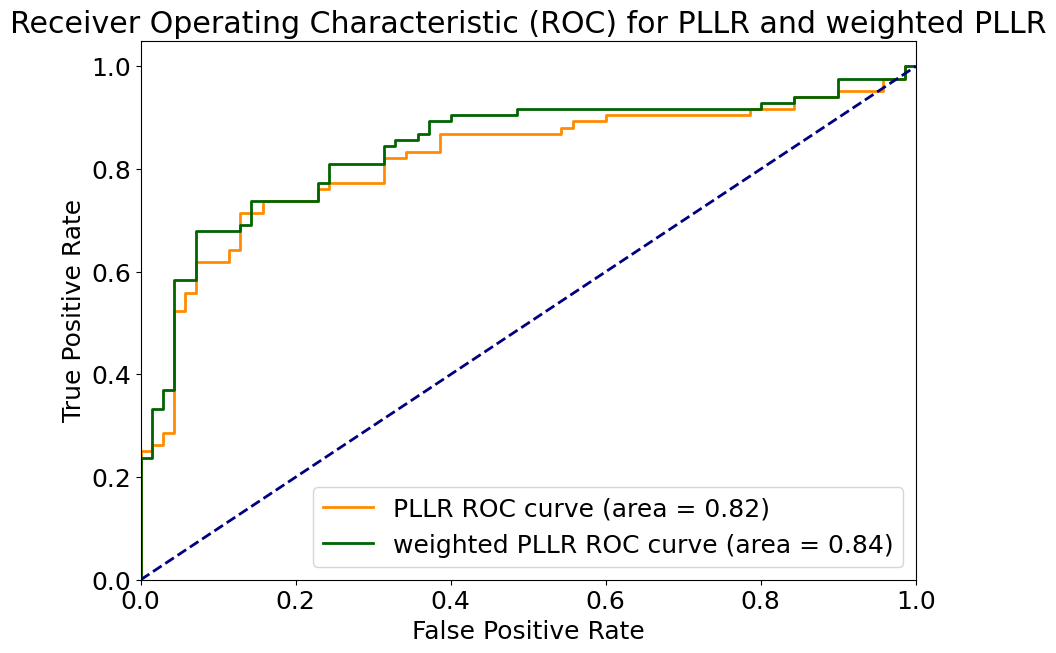}
        \caption{ROC for pretrained ESM2 on ARM}
        \label{fig:zs-arm-c}
    \end{subfigure}
    \caption{Zero-shot performances on ARM}
    \label{fig:zs-arm-m}
\end{figure}

Similar to our previous observations on CM, Figure~\ref{fig:ft-arm-m} and Figure~\ref{fig:ft-arm} (in subsection~\ref{sp:pr-ft-arm} of the supplementary material) show the ROC and PR curves for both fine-tuned ESM1b and ESM2, respectively. The fine-tuned ESM1b achieves an AUC of $0.94$ and an AUPR of $0.95$ for weighted PLLR on ARM in Figure~\ref{fig:ft-arm-a} and~\ref{fig:ft-arm-b}. The fine-tuned ESM2 also achieves a slight performance increase; for instance, the AUPR for the weighted PLLR rises from $0.88$ in Figure~\ref{fig:zs-arm-d} to $0.9$ in Figure~\ref{fig:ft-arm-d}. This demonstrates that our ProPath more effectively captures the PLLR in rare missense variants for variant pathogenicity in ARM.
\begin{figure}[!htbp]
    \centering
    \begin{subfigure}{0.45\textwidth}
        \centering
        \includegraphics[width=7.5cm,height=5cm]{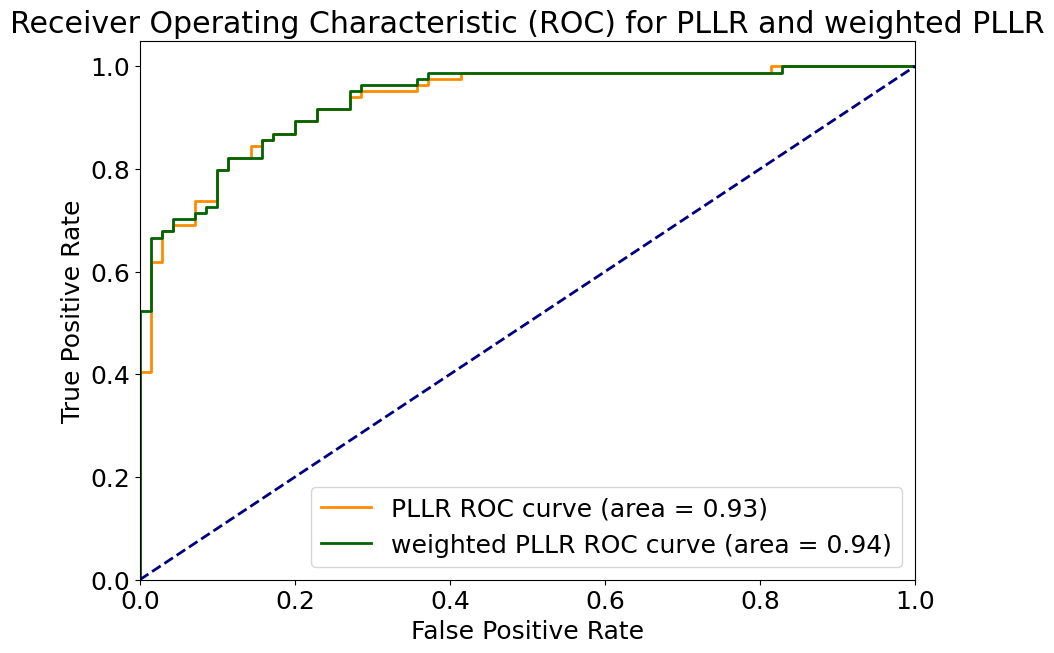}
        \caption{ROC for fine-tuned ProPath via ESM1b on ARM}
        \label{fig:ft-arm-a}
    \end{subfigure}
    \begin{subfigure}{0.45\textwidth}
        \centering
        \includegraphics[width=7.5cm,height=5cm]{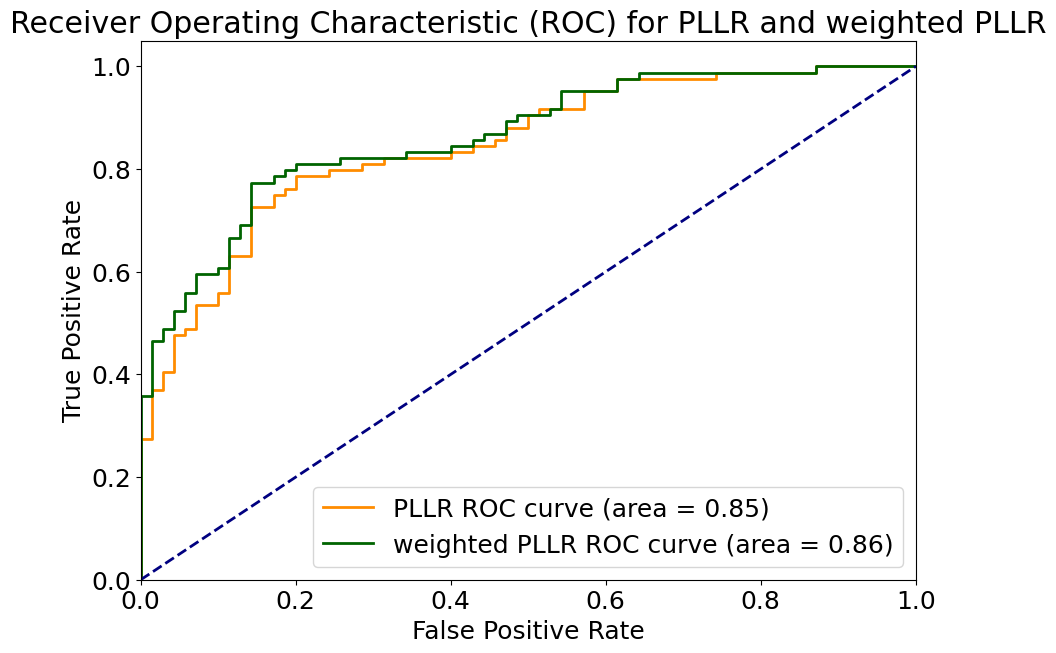}
        \caption{ROC for fine-tuned ProPath via ESM2 on ARM}
        \label{fig:ft-arm-c}
    \end{subfigure}
    \caption{Fine-tuned ProPath performances on ARM.}
    \label{fig:ft-arm-m}
\end{figure}

\paragraph{ProPath outperforms all baseline methods on ARM.}
%We further evaluate the ability of our DS-SPLM to capture disease-specific information related to ARM via comparing with the aforementioned five baseline methods shown in Table~\ref{tab:arm-baselines}. The best performances in terms of AUC and AUPR are shown in bold. Our DS-SPLM, utilizing ESM1b as its foundational model, outperforms the five baselines with an AUC of $0.94$ and an AUPR of $0.95$. This result outperforms the second best result AdaBoost (shown in italic) with a $5\%$ AUC increase. Therefore, this reaffirms the capability of DS-SPLM as a robust tool for disease-specific VEP in discerning disease associations for both datasets.

Finally, we compare ProPath performances in ARM with the five benchmark methods detailed in Table~\ref{tab:arm-baselines}, where the best AUC and AUPR are highlighted in bold. With ESM1b as its base, our ProPath surpasses all baselines, recording an AUC of $0.94$ and an AUPR of $0.95$. This performance marks a $5\%$ AUC improvement over the second-best performer, AdaBoost (indicated in italic). Such finding further shows ProPath sets a new benchmark by obtaining the highest statistics across all baselines in both CM and ARM datasets.

\begin{table}[!htbp]
\caption{ Performance for arrhythmias variant pathogenicity prediction.}
\label{tab:arm-baselines}
\centering
\scalebox{1}{
 \begin{tabular}{l |>{\centering\arraybackslash}p{5cm} >{\centering\arraybackslash}p{1.5cm} >{\centering\arraybackslash}p{1.5cm} >{\centering\arraybackslash}p{1.5cm} }
    \hline
      \multirow{2}{*}{\textbf{Category}}   & \multirow{2}{*}{\textbf{Algorithm}}  & \textbf{Disease-specific} & \multirow{2}{*}{\textbf{AUC}} & \multirow{2}{*}{\textbf{AUPR}}\\
      \hline
    \multirow{3}{*}{Boosting models} 
         & XGBoost~\citep{zhang2021disease} & $\checkmark$& $0.90$ & $0.88$  \\
         & AdaBoost~\citep{zhang2021disease}& $\checkmark$ & $\textit{0.90}$ & $\textit{0.90}$ \\
         & M-CAP~\citep{jagadeesh2016m} & $\times$ & $0.85$ & $0.81$\\
          \hline
    \multirow{1}{*}{Ensemble models}  
         & REVEL~\citep{ioannidis2016revel} & $\times$ & $0.81$ & $0.79$ \\
         \hline
    \multirow{2}{*}{Language models}
         & ESM1b~\citep{brandes2023genome} & $\times$& $0.90$ & $0.89$ \\
         & ProPath  &$\checkmark$ & $\mathbf{0.94}$ & $\mathbf{0.95}$ \\
         \hline
    \end{tabular}
}
\end{table}
\subsection{Distribution of PLLR values shows ProPath effectively identifies pathogenic and benign rare missense genetic variants.}
\begin{wrapfigure}[13]{r}{0.7\textwidth}
%\begin{figure}
    \centering
    \includegraphics[width=10cm,height=5cm]{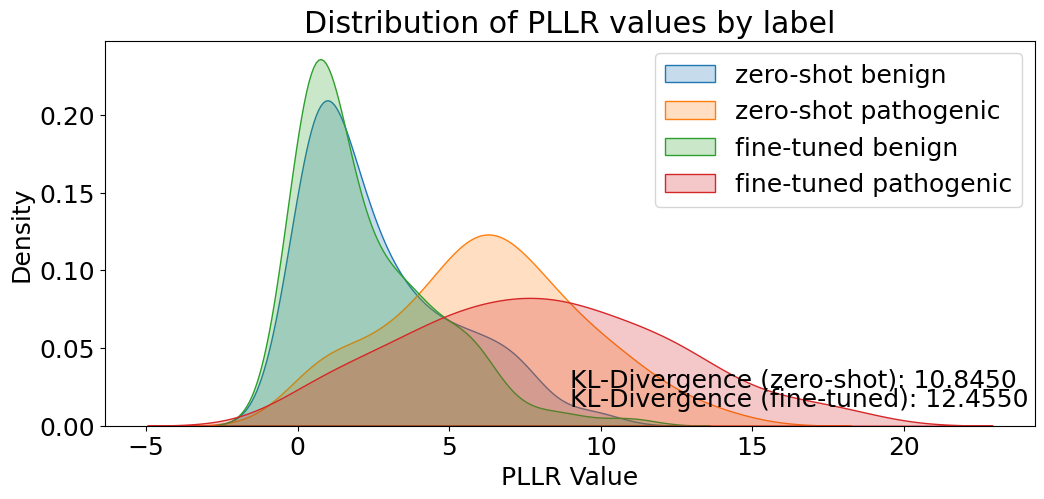}
    \caption{Distribution of PLLR values for benign and pathogenic sequences in both zero-shot and fine-tuned scenarios.}
    \label{fig:distribution}
\end{wrapfigure}

In Figure~\ref{fig:distribution}, we show the distribution of PLLR values for benign and pathogenic sequences in both zero-shot and fine-tuned scenarios on CM. We observe that the zero-shot model computes the PLLR difference between benign and pathogenic sequences with a smaller KL-divergence value, specifically $10.8450$. After fine-tuning, the PLLR for the benign data is closer to zero, while the PLLR for the pathogenic sequences becomes larger. This indicates that our ProPath effectively learned disease-specific information to identify pathogenic and benign rare missense genetic variants in cardiomyopathies.

\subsection{Fine-tuned MLM logits capture disease-specific intolerance for pathogenic variants.}
\begin{figure}[!htbp]
    \centering
    \begin{subfigure}{1\textwidth}
    \hspace{0.5cm}
        \includegraphics[width=17cm,height=4.6cm]{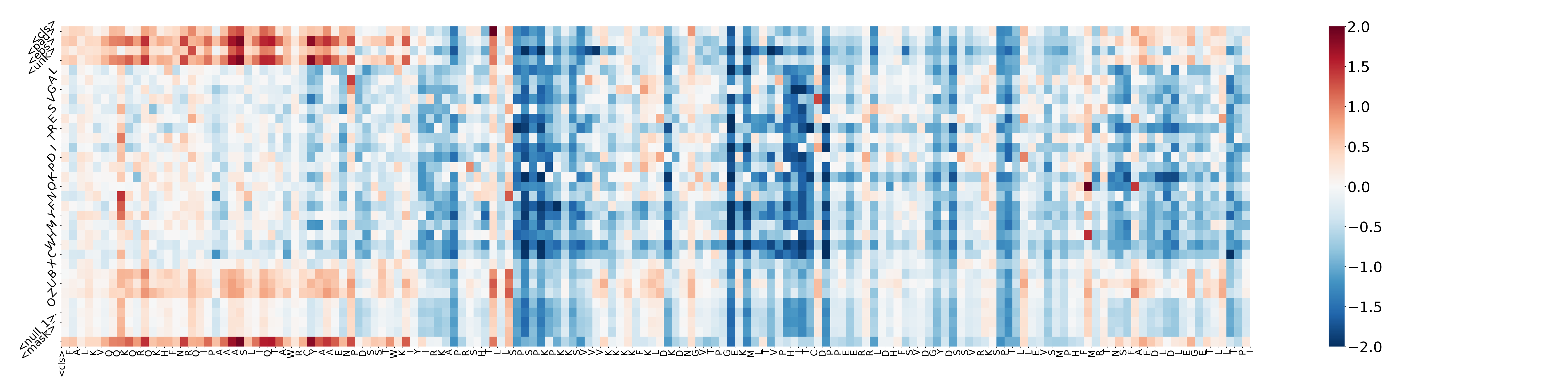}
        \caption{The MLM logits differences between zero-shot ESM1b and ProPath fine-tuned on CM.}
        \label{fig:}
    \end{subfigure}

    \begin{subfigure}{1\textwidth}
    \hspace{0.5cm}
        \includegraphics[width=17cm,height=4.6cm]{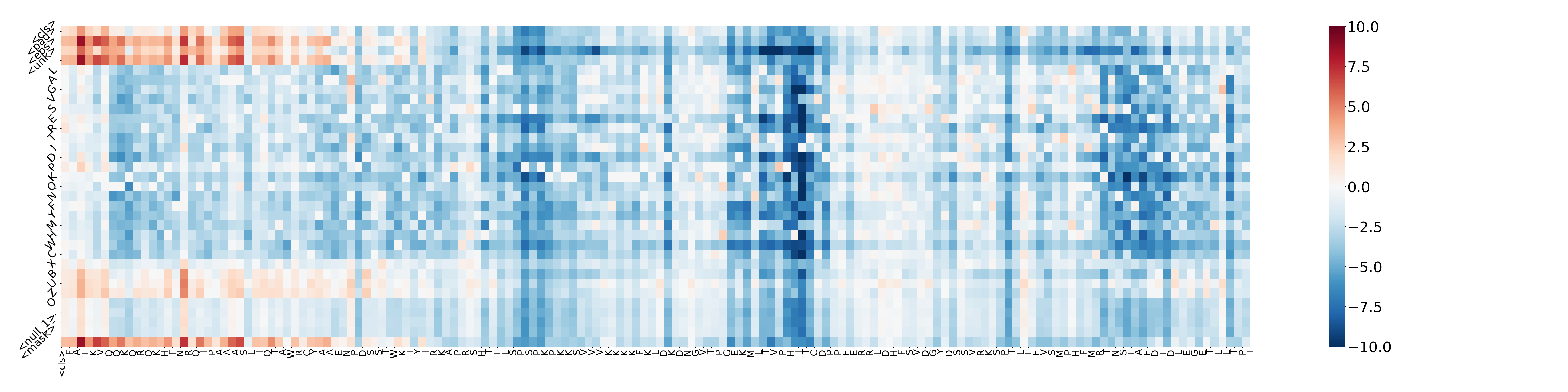}
        \caption{The MLM logits differences between zero-shot ESM1b and ProPath fine-tuned on ARM.}
        \label{fig:}
    \end{subfigure}
    \caption{The MLM logits differences between zero-shot ESM1b and fine-tuned ProPath. X-axis is the protein sequence, and y-axis is the dictionary.}
    \label{fig:mlm_logits}
\end{figure}

Given the improved VEP performance after disease-specific fine-tune in our ProPath, we sought to uncover the positional basis of disease-specific pathogenic variant intolerance, and evaluate the similarity and difference of such variant intolerance between CM and ARM at single amino acid resolution. Because zero-shot ESM-1b performed subpar on both the CM and ARM variant sets, we reasoned that ProPath's superior performance is a result of more accurate pathogenic variant intolerance maps that are specific to CM and ARM, respectively. To quantify the positional impacts after fine-tuning, we calculated the MLM logits difference between the zero-shot MLM logits and the fine-tuned MLM logits in a representative wild-type protein sequence in one of the cardiovascular disease-relevant genes. We performed this analysis on the same protein sequence but separately on ProPath fine-tuned CM and ARM models (Fig. \ref{fig:mlm_logits}).

Intuitively, as MLM logits represent how likely each amino acid could replace the masked residue, a shift in logits after fine-tuning demonstrates intolerance for disease-specific pathogenic variants (blue in Fig. \ref{fig:mlm_logits}). Since CM and ARM are two closely related cardiovascular conditions, we observed an overall similar pattern of intolerance between CM and ARM (blue bands in Fig. \ref{fig:mlm_logits}). Upon a closer comparison, many fine-grained differences are evident: CM is more tolerant towards missense variants in the beginning part of the protein compared to ARM, while ARM has a more profound MLM logits difference in scale. This suggests our ProPath successfully captured the disease-specific intolerance for pathogenic variant at single amino-acid resolution. Future functional studies will reveal if such patterns are consistent with protein domain annotations and our predictions can be validated in cohort-based experimental designs.

%\section{Relates Works}
\section{Conclusion}
Conventional unsupervised VEP methods fall short in being disease-specific, posing significant limitations in their clinical adaptability and relevance at the point-of-care. Such methods produce a score for each missense genetic variant, indicating potential damage to the protein's function or structure related to the disease. However, they don't factor in the intricate, disease-specific parameters that dictate the gene-disease relationship. Addressing this gap, we introduce a disease-specific \textsc{pro}tein language model for variant \textsc{path}ogenicity, i.e., ProPath, designed for a fine-grained capture of the pseudo-log-likelihood ratio in rare missense variants. In our evaluation using clinical variant sets from inherited cardiomyopathies and arrhythmias, ProPath's performance outperforms pre-trained language models. For instance, it outperforms the pre-trained ESM1b with an over $5\%$ improvement in AUC across both cardiomyopathy and arrhythmia datasets. In summary, ProPath provides a powerful tool for disease-specific variant effect prediction, enhancing the understanding of disease associations and offering significant clinical value.

\medskip

{
%\small

\bibliographystyle{apalike}
\bibliography{sample}
}

\newpage
\section{Supplementary Material}
\setcounter{table}{-1} % to start with 0
\counterwithin*{table}{section} % to remove any section dependency if you're numbering within sections
\renewcommand{\thetable}{S\arabic{table}} % to prepend "S" to the table number

\setcounter{figure}{-1} % to start with 0
\counterwithin*{figure}{section} % to remove any section dependency if you're numbering within sections
\renewcommand{\thefigure}{S\arabic{figure}} % to prepend "S" to the figure number

\subsection{MLM logtis for $s^{\text{WT}}$ and $s^{\text{mut}}$}\label{sec:mlm_logits}
\begin{figure}[!htbp]
    \centering
    \begin{subfigure}{0.45\textwidth}
        \centering
        \includegraphics[width=7.5cm,height=5cm]{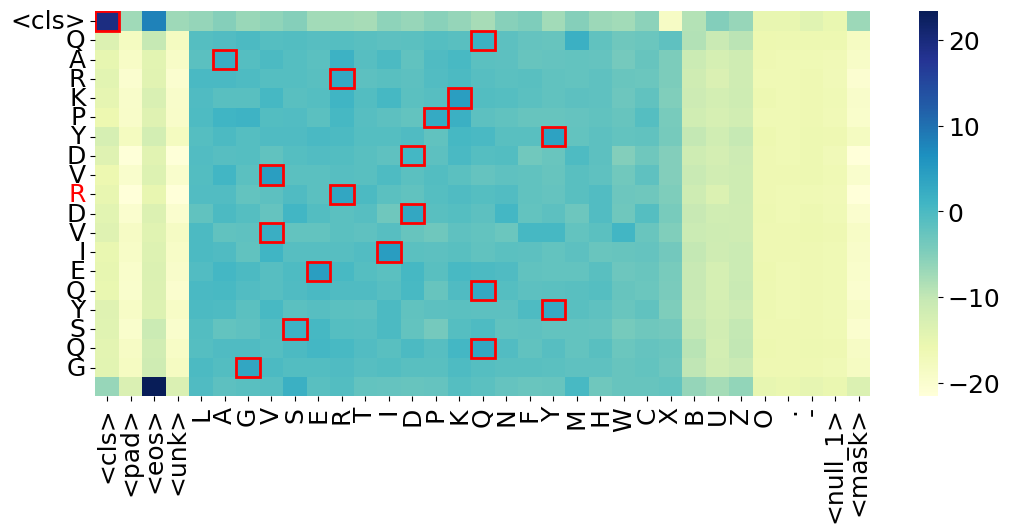}
        \caption{MLM logtis for $s^{\text{WT}}$}
        \label{fig:mlm-1}
    \end{subfigure}
    \hspace{1pt}
    \begin{subfigure}{0.45\textwidth}
        \centering
        \includegraphics[width=7cm,height=5cm]{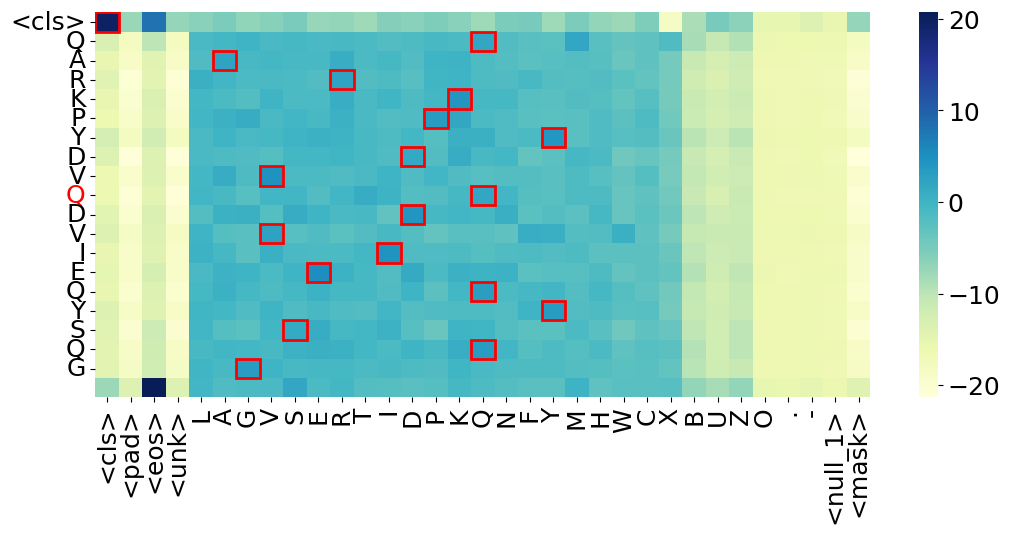}
        \caption{MLM logtis for $s^{\text{mut}}$}
        \label{fig:mlm-2}
    \end{subfigure}
    \caption{MLM logtis for $s^{\text{WT}}$ and $s^{\text{mut}}$, where the x-axis denotes the dictionary and the y-axis denotes the sequences, e.g., $s^{\text{WT}}$ and $s^{\text{mut}}$. The red frames emphasize the likelihoods, as predicted by the protein language model, for the occurrence of amino acid $s_i$ at position $i$ within the sequence $s$.}
    \label{fig:mlm}
\end{figure}

\subsection{Details for dataset and set-up}\label{sp:dataset}

In our experiments, we focus on clinical variant sets pertaining to inherited cardiomyopathies and arrhythmias. We employed a pre-compiled dataset of rare missense pathogenic and benign variants, defined by a cohort-based approach, in cardiomyopathy and arrhythmias, respectively. The details can be found in the previous report ~\citet{zhang2021disease}. Briefly, the authors collected 356 unique rare missense variants specifically associated with cardiomyopathy genes. The authors curated a set of 252 unique rare missense variants linked with established arrhythmia genes from ClinVar. For a benign set, they gathered 302 unique rare missense variants in cardiomyopathy genes and 237 in arrhythmia genes from the targeted sequencing of 2,090 healthy individuals. For training and testing set distribution, 440 and 326 cardiomyopathy variants were designated for training and testing, respectively. For the arrhythmias category, 218 variants were used for training and 166 for testing. The statistics for both datasets are shown in Table~\ref{tab:dataset}.
\begin{table}[!htbp]
    \caption{Cardiomyopathies and arrhythmias datasets}
    \label{tab:dataset}
    \centering
    \begin{tabular}{l|c c c c c c}
    \hline
      & \multicolumn{3}{c}{Cardiomyopathies} & \multicolumn{3}{c}{Arrhythmias}\\
     & Pathogenic & Benign & Total & Pathogenic & Benign & Total \\
    \hline
     Training    & 238 & 202 & 440 & 168 & 158 & 326 \\
      Test    & 118  & 100 & 218 & 84 & 79 & 163\\
      Total & 356 & 302 & 658 & 252 & 237 & 489\\
      \hline
    \end{tabular}
\end{table}

For fine-tuning on the ProPath, we use a batch size of $8$; the evaluation batch size is also set to $8$. The model is trained over $10$ epochs on all datasets for both tasks. We employ the Adam optimizer, with a learning rate of $1\text{e}-5$. Additionally, we implement a warmup ratio of $0.1$ followed by linear decay. The L2 regularization weight decay is set at $0.01$. For sequences longer than 1024 that exceed the max length limit of protein language models, we truncate them to a length of 1024, centering them around the variant position. We fine-tune our ProPath on two base protein language models, i.e., esm1b\_t33\_650M\_UR50S (ESM1b)~\citep{rives2021biological} and esm2\_t33\_650M\_UR50D (ESM2)~\citep{lin2023evolutionary}. ESM1b is a protein language model endowed with $650$ million parameters and trained on a corpus of $250$ million protein sequences spanning various organisms. Its training involved the MLM task, where certain residues from input sequences are randomly obscured, challenging the model to accurately predict the correct amino acid for each masked position. On the other hand, ESM2 the latest model in the ESM family with improvements in architectures and training parameters, and is benchmarked to outperform ESM1b at a comparable number of parameters. Over the course of training, ESM2 sees around 65 million unique sequences.

\subsection{PR for pretrained protein language models on CM}\label{sp:pr-zs-cm}
\begin{figure}[!htbp]
    \centering
    \begin{subfigure}{0.45\textwidth}
        \centering
        \includegraphics[width=7cm,height=5cm]{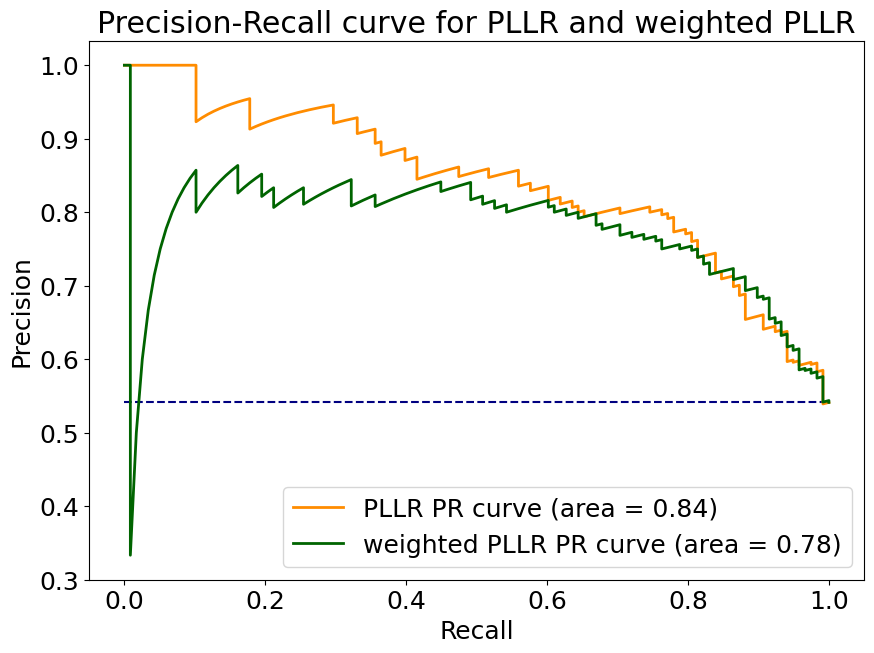}
        \caption{PR for pretrained ESM1b on CM}
        \label{fig:zs-cm-b}
    \end{subfigure}
    \begin{subfigure}{0.45\textwidth}
        \centering
        \includegraphics[width=7cm,height=5cm]{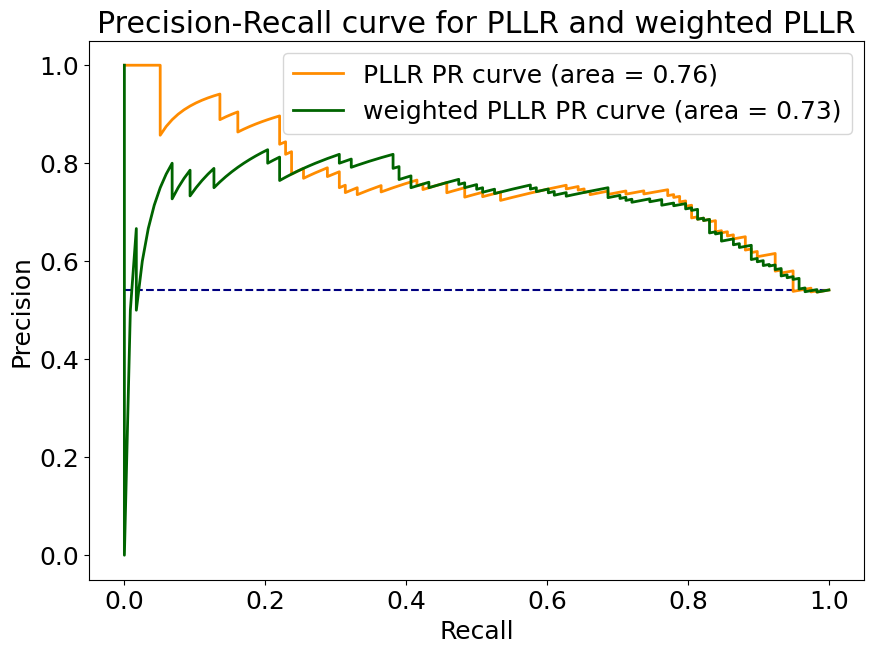}
        \caption{PR for pretrained ESM2 on CM}
        \label{fig:zs-cm-d}
    \end{subfigure}
   \caption{Zero-shot performances on CM.}
    \label{fig:zs-cm}
\end{figure}

\subsection{PR for fine-tuned ProPath on CM}\label{sp:pr-ft-cm}
\begin{figure}[!htbp]
    \centering
    \begin{subfigure}{0.45\textwidth}
        \centering
        \includegraphics[width=7cm,height=5cm]{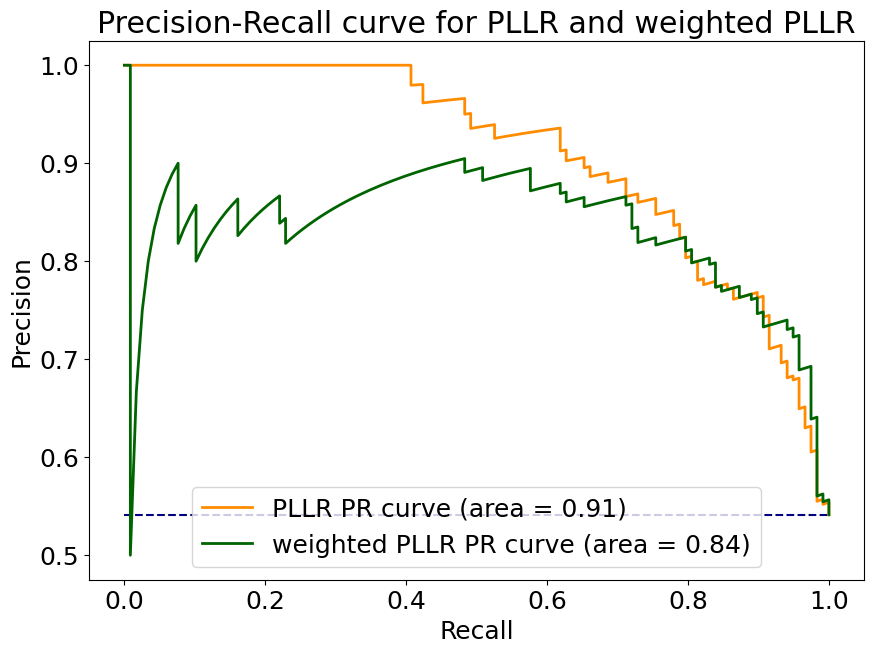}
        \caption{PR for fine-tuned ProPath via ESM1b on CM}
        \label{fig:peft}
    \end{subfigure}
    \begin{subfigure}{0.45\textwidth}
        \centering
        \includegraphics[width=7cm,height=5cm]{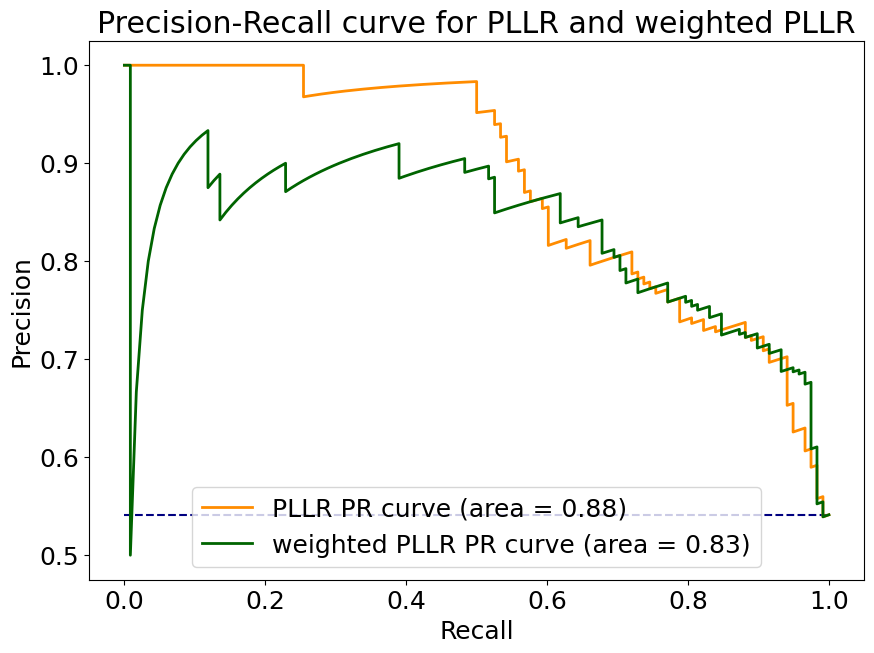}
        \caption{PR for fine-tuned ProPath via ESM2 on CM}
        \label{fig:}
    \end{subfigure}
    \caption{Fine-tuned ProPath performances on CM.}
    \label{fig:fs-cm}
\end{figure}
\newpage
\subsection{PR for pretrained protein language models on ARM}\label{sp:pr-zs-arm}
\begin{figure}[!htbp]
    \centering
    \begin{subfigure}{0.45\textwidth}
        \centering
        \includegraphics[width=7cm,height=5cm]{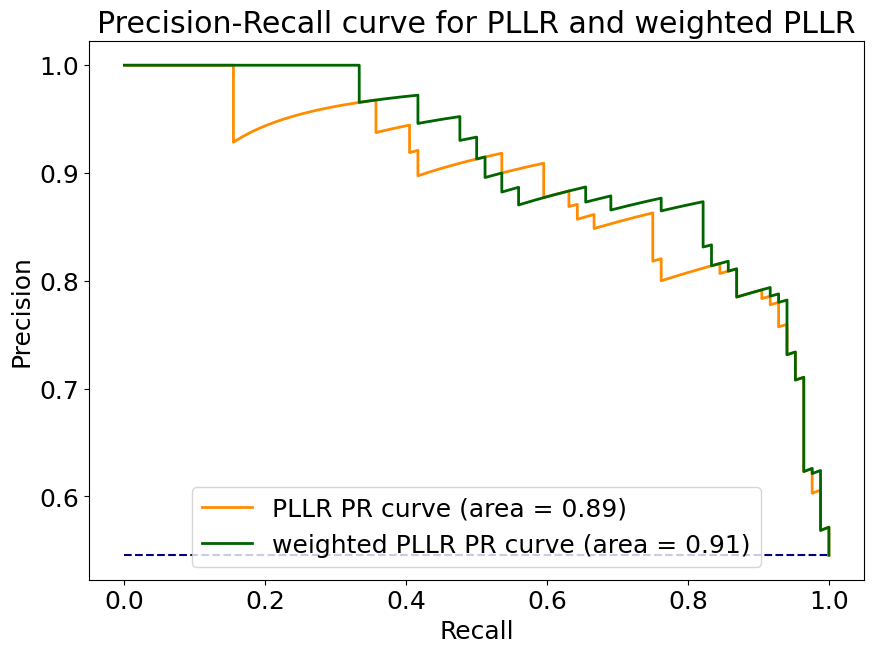}
        \caption{PR for pretrained ESM1b on ARM}
        \label{fig:zs-arm-b}
    \end{subfigure}
    \begin{subfigure}{0.45\textwidth}
        \centering
        \includegraphics[width=7cm,height=5cm]{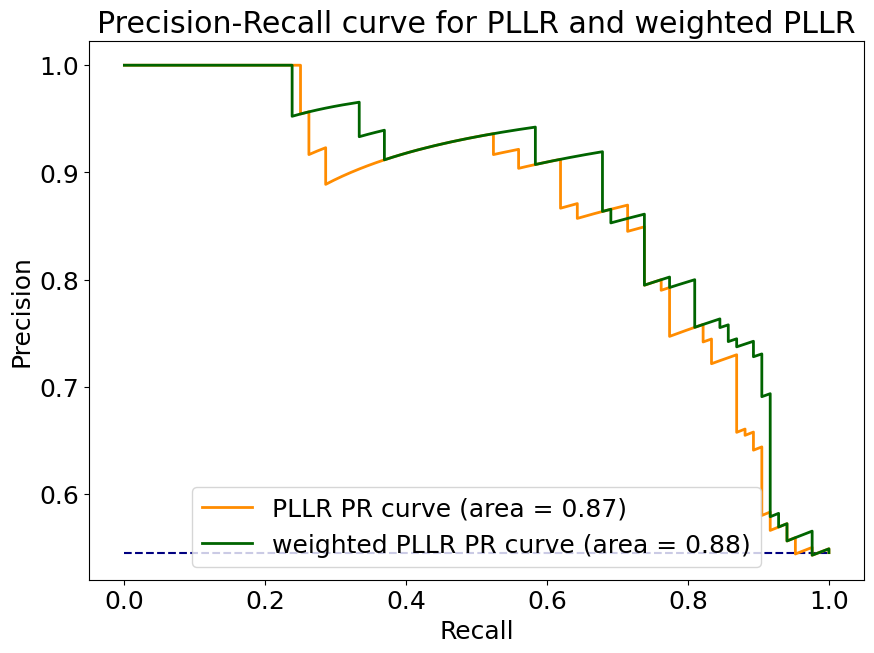}
        \caption{PR for pretrained ESM2 on ARM}
        \label{fig:zs-arm-d}
    \end{subfigure}
    \caption{Zero-shot performances on ARM}
    \label{fig:zs-arm}
\end{figure}

\subsection{PR for fine-tuned ProPath on ARM}\label{sp:pr-ft-arm}
\begin{figure}[!htbp]
    \centering
    \begin{subfigure}{0.45\textwidth}
        \centering
        \includegraphics[width=7cm,height=5cm]{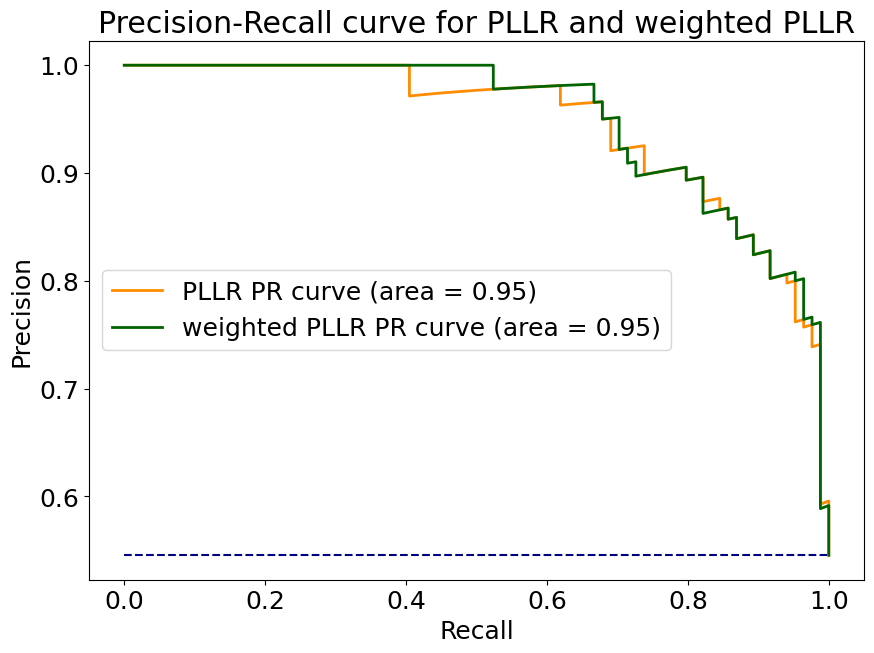}
        \caption{PR for fine-tuned ProPath via ESM1b on ARM}
        \label{fig:ft-arm-b}
    \end{subfigure}
    \begin{subfigure}{0.45\textwidth}
        \centering
        \includegraphics[width=7cm,height=5cm]{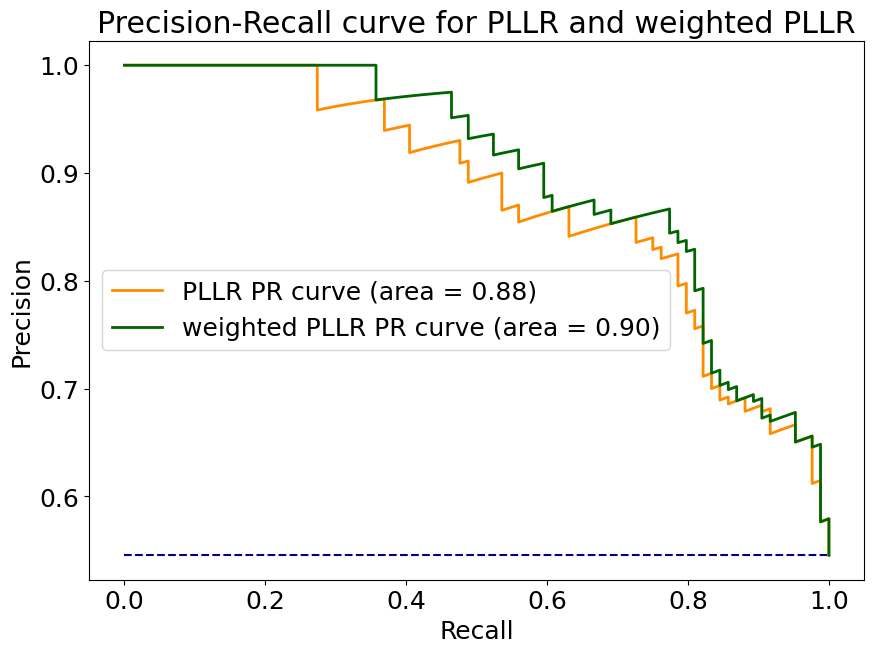}
        \caption{PR for fine-tuned ProPath via ESM2 on ARM}
        \label{fig:ft-arm-d}
    \end{subfigure}
    \caption{Fine-tuned ProPath performances on ARM.}
    \label{fig:ft-arm}
\end{figure}

%%%%%%%%%%%%%%%%%%%%%%%%%%%%%%%%%%%%%%%%%%%%%%%%%%%%%%%%%%%%

\end{document}